# Cosmological Properties of Eternally Collapsing Objects (ECOs)

© Abhas Mitra

Theoretical Astrophysics Section, BARC, Mumbai -400085, India
Email: amitra@barc.gov.in

**Abstract:** We show that the integration constant in the vacuum Schwarzschild solution has the unique value, $\alpha_0=0$, accordingly Black Holes too have the same unique value of mass $M_0=0$. Therefore the so-called Black Holes Candidates (BHC) cannot be true BHs. It is also shown that continued collapse of sufficiently massive bodies would generate radiation pressure and energy dominated quasistatic objects having surface gravitational redshifts $z\gg 1$. Under the assumption of baryon number conservation, such objects would take infinite time to collapse to the idealized BH state with $M=0$ and $z=\infty$. The local temperature of such a stellar mass Eternally Collapsing Object (ECO) would be above Quark Gluon Phase transition. ECOs would undergo intermittent violent radiative eruptions and pollute the interstellar medium with freshly made hydrogen out of their QGP and also the light elements cooked in their envelope. It is shown that the extremely redshifted observed temperature of an ECO could be 2.75 K and superposition of ECO background radiation might generate the microwave back ground radiation. The predicted 2.75 K luminosity for the galactic centre ECO, i.e., Sgr A$^*$, L~$3\times 10^{36}$ erg/s , nicely matches with the corresponding estimate by Wilkinson Microwave Anisotropy Probe (WMAP).

## 1. Introduction: Black Holes?

There are compact objects with several solar masses ($M_*$) in many X-ray binaries and we are also aware of compact objects of even billion stellar masses in the core of some Active Galactic Nuclei. Since it is known that, the masses of *cold and strictly static* objects cannot be much higher than $3 M_*$, it is generally believed that, such objects are ``Black Holes'' because it is (erroneously) believed that all compact objects with masses $> 3M_*$ must be BHs and nothing else. This is so because the celebrated vacuum Schwarzschild solution

$$ds^2 = -(1-\alpha_0/R)dT^2 + (1-\alpha_0/R)^{-1}dR^2 + R^2(d\theta^2 + \sin^2\theta d\phi^2), \tag{1}$$

where $\alpha_0 = 2GM_0/c^2$, G is gravitational constant and c is the speed of light, apparently suggests that the parameter $\alpha_0$ could be arbitrarily high. Here θ and φ are the usual polar coordinates, ``T'' is the time and ``R'' is the (invariant) circumference coordinate. Recall that $\alpha_0$ arose in this equation as an ``integration constant'' and we will show shortly that the *value of this integration constant is unique, $\alpha_0=0$*. Therefore the so-called BHCs (or anything else) with finite masses cannot be true BHs. It may be also noted that there is really no upper mass limit on sufficiently ``hot'' objects and even Newtonian gravity allows for existence of radiation supported quasistatic ``supermassive'' stars. Since all hot objects must radiate, they cannot be described by any strictly *static* solution such as the Schwarzschild solution. This should hint that, the massive BHCs could actually be extremely hot radiating quasistatic objects rather than cold and strictly static objects.

It will be also shown that no trapped surface is formed in General Relativistic (GR) gravitational collapse:

$$\frac{2GM(r,t)}{Rc^2} \le 1 \tag{2}$$

where ``r'' is the comoving radial coordinate characterizing a given mass shell of a spherical fluid and ``t'' is comoving time. Eq.(2) is in accordance with the fact that there cannot be any finite mass BH (or naked singularity) and the only BH mass which is permitted is $M_0 = 0$. Next we shall argue that GR collapse of sufficiently massive or dense objects should indeed result in the formation of extremely hot quasistatic objects with appropriate high values of z. This is so because for high z, the collapse generated heat/radiation must get trapped by self-gravity. Then, sooner or later, the trapped radiation pressure must attain the critical Eddington value at which, by definition, the outward radiation force = Inward gravitational pull. Although an ECO is extremely hot locally, its temperature seen by a distant observer must be extremely low because of its extremely high value of gravitational redshift z. It will be found that the observed temperature of an ECO is independent of its mass and can very well 2.75 K. If so, the back ground microwave radiation could result from the contribution of ECOs in the observable universe.

The local temperatures of stellar mass ECOs are found to be well above 100 MeV and therefore such ECOs are made of QGP like the mythical ``hot early universe'' in Big Bang cosmology. The ECO envelope/photosphere would be relatively cooler and QGP would condense to form neutrons and protons. Further, in regions of temperature ~ 1 MeV light elements would be cooked out of these neutrons and protons as it is supposed to happen in Big-Bang cosmology. The cooked, hydrogen and light isotopes would remain trapped by strong self-gravity for a quiescent ECO. But an ECO being supported entirely by radiation pressure is extremely vulnerable to radiation driven eruptions. During such eruptions, an ECO would gorge out not only light isotopes but pure hydrogen too in the interstellar medium (ISM). At the same time an ECO is accreting from the ISM. Despite such accretions, and ECO may get evaporated completely because of unending radiative eruptions.

## 2. The Unique Value of Black Hole Mass

The extended Eddington-Finkelstein metric which describes both interior and exterior spacetimes of a BH is

$$ds^2 = -(1-\alpha_0/R)dT_*^2 + (1-\alpha_0/R)^{-1}dR^2 \mp (2\alpha_0/R)dT_*dR + R^2(d\theta^2 + \sin^2\theta d\phi^2) \tag{3}$$

where the Finkelstein coordinates are

$$T_* = T \mp \log\left(\frac{R-\alpha_0}{\alpha_0}\right);\ldots R_* = R,\ldots \theta_* = \theta,\ldots \phi_* = \phi \tag{4}$$

The corresponding metric coefficients are

$$g_{T_*T_*} = -(1-\alpha_0/R),\ldots g_{R_*R_*} = (1+\alpha_0/R)^{-1},\ldots g_{T_*R_*} = g_{R_*T_*} = \alpha_0/R \tag{5}$$

The determinant for both the metrics are same:

$$g_* = -g_{\theta\theta}g_{\phi\phi}(g^2_{T_*R_*} - g_{T_*T_*}g_{R_*R_*}) = -R^4\sin^2\theta = g \tag{6}$$

Now let us apply the principle of *invariance of 4-volume* for the two coordinate systems (T, R, θ, φ) and (T$_*$, R$_*$, θ$_*$, φ$_*$):

$$\iiiint \sqrt{-g_*}\, dT_* dR_* d\theta_* d\phi_* = \iiiint \sqrt{-g}\, dT dR d\theta d\phi \tag{7}$$

Since $g = g_* = -R^4\sin^2\theta$, we can write the foregoing equation as

$$\iiiint R^2 \sin\theta\, dT_* dR\, d\theta\, d\phi = \iiiint R^2 \sin\theta\, dT dR d\theta d\phi \tag{8}$$

The integration over the angular coordinates can be easily carried out and cancelled from both sides. Then one will be left with

$$\iint R^2 dT dR \mp \alpha_0 \iint \frac{R^2}{R-\alpha_0} dR dR = \iint R^2 dT dR \tag{9}$$

where we have used the fact that

$$dT_* = dT \mp \frac{\alpha_0}{R-\alpha_0} dR \tag{10}$$

From, Eq.(9), one has

$$\alpha_0 \iint \frac{R^2}{R-\alpha_0} dR dR = 0 \tag{11}$$

And this equation can be satisfied only if $\alpha_0 \equiv 0, i.e., M_0 \equiv 0$. Thus BHs have a unique mass M=0. and the so-called BHCs are not true BHs [1,2,3].

## 3. Trapped Surfaces?

Let us consider the metric for spherically symmetric fluid in commoving coordinates:
$$ds^2 = g_{00} dt^2 + g_{rr} dr^2 + R^2 (d\theta^2 + \sin^2\theta d\phi^2) \tag{12}$$
For radial motion, $d\theta = d\phi = 0$ and the metric becomes
$$ds^2 = g_{00} dt^2 (1 - x^2) \tag{13}$$
where the auxiliary parameter

$$x = \frac{\sqrt{-g_{rr}} dr}{\sqrt{g_{00}} dt} \tag{14}$$

Eq.(13) may be rewritten as
$$(1 - x^2) = \frac{1}{g_{00}} \frac{ds^2}{dt^2} \tag{15}$$

Suppose an observer sitting at a fixed R is observing the fluid as it passes by. For such a $R = cons$ observer, we will have
$$dR(r,t) = 0 = \dot{R} dt + R' dr \tag{16}$$
where a dot denotes partial differentiation w.r.t. t and a prime denotes the same w.r.t. r. So, at a fixed R:

$$\frac{dr}{dt} = -\frac{\dot{R}}{R'} \tag{17}$$

and the corresponding x is
$$x = x_c = \frac{\sqrt{-g_{rr}} dr}{\sqrt{g_{00}} dt} = -\frac{\sqrt{-g_{rr}} \dot{R}}{\sqrt{g_{00}} R'} \tag{18}$$
Now let us define:
$$\Gamma = \frac{R'}{\sqrt{-g_{rr}}}, \ldots\ldots U = \frac{\dot{R}}{\sqrt{g_{00}}} \tag{19}$$
so that Eq.(18) yields
$$x_c = \frac{-U}{\Gamma}; \ldots U = -x_c \Gamma \tag{20}$$

The gravitational mass of the collapsing fluid is defined through the following equation:
$$\Gamma^2 = 1 + U^2 - \frac{2GM(r,t)}{Rc^2} \tag{21}$$
Using Eq.(20) in (21) and transposing, we have

$$\Gamma^2 (1 - x^2_c) = 1 - \frac{2GM(r,t)}{Rc^2} \tag{22}$$

By using Eqs. (15) and (20) in the foregoing eq., we find

$$\frac{R'^2}{-g_{rr} g_{00}} \frac{ds^2}{dt^2} = 1 - \frac{2GM(r,t)}{Rc^2} \tag{23}$$

Since the determinant of the metric tensor $g = R^4 \sin^2 \theta g_{00} g_{rr} \leq 0$, we must have $-g_{00} g_{rr} \geq 0$. Then Eq.(23) shows that its LHS is positive. So must be its RHS, and this implies that

$$\frac{2M(r,t)}{R} \leq 1 \tag{24}$$

Thus, in a most general manner, it is found that *there cannot be any trapped surface ever*[4-7]

## 4. Eternally Collapsing Objects

Absence of trapped surfaces means that there cannot be any singularity in GR collapse and thus, *in a strict sense,* the collapse must proceed indefinitely not only w.r.t. a faraway observer but also w.r.t. any observer and in particular, the comoving observer. Thus GR continued collapse solution cannot result in the formation of any strictly static cold object. Eq.(24), however, as a limiting case, allows for the formation of a BH of mass M=0 as R→ 0. Since a zero mass BH has got no energy to radiate, the final state of GR continued collapse, under the assumption of baryon number conservation, is indeed a zero mass BH in accordance with the result $\alpha_0 = 2GM_0/c^2 \equiv 0$.

How can such a state be physically achieved? It is known that whether it is Newtonian or GR regime gravitational collapse /contraction must result in emission of radiation (otherwise there would be no collapse/ contraction) [8,9]. So mass of the collapsing fluid must keep on diminishing and should asymptotically lead to M→0 BH state. Irrespective of the question of uniqueness of mass, a BH Event Horizon has a radius

$$R = R_s = \alpha_0 = \frac{2GM}{c^2} \tag{25}$$

and this corresponds to a surface of gravitational redshift $z = \infty$ because one defines

$$z = (1 - 2GM/Rc^2)^{-1/2} - 1 \tag{26}$$

Thus by definition, in order to become a BH, the collapsing object must pass through stages characterized by z>>1. Let us probe the consequence of such intermediate stages for the radiation emitted during the collapse. As long as $z < \sqrt{3} - 1$, the emitted quanta would manage to avoid entrapment and move away to infinity. But when the body would be so compact as to lie within the ``photon sphere'', i.e., $R < (3/2)R_s, or,..z > \sqrt{3} - 1$, the radiation emitted only within a cone defined by semi-angle $\theta_c$ :

$$\sin\theta_c = \frac{\sqrt{27}}{2}(1 - R_s/R)^{1/2}(R_s/R) \tag{27}$$

will be able to eventually escape. Radiation emitted in the rest of the hemisphere would return within the compact body. In fact, it can be seen that, this gravitational trapping of the quanta is synonymous with the fact that they move in closed circular orbits for $z > \sqrt{3} - 1$. At large z, $R \approx R_s$ and from Eq.(27), one can see that

$$\sin\theta_c \to \theta_c \approx \frac{\sqrt{27}}{2}(1+z)^{-1};..z >> 1 \tag{28}$$

The solid angle formed by the escaping radiation is

$$\Omega_c \approx \pi\theta_c^2 \approx \frac{27\pi}{4}(1+z)^{-2} \tag{29}$$

The chance of escape of radiation therefore decreases as

$$\frac{\Omega_c}{2\pi} \propto (1+z)^{-2} \tag{30}$$

In the absence of any self-gravitational trapping, the outward heat/radiation flux (q) would increase as ~$R^{-3}$ during collapse. And when we incorporate the relativistic effect of gravitational trapping, we will have

$$q_{trap} \sim R^{-3}(1+z)^2 \tag{31}$$

4.1: Eddington Luminosity:

In GR, the locally defined Eddington luminosity for a compact object of mass M is

$$L_{ed} = \frac{4\pi GMc}{\kappa}(1+z) \qquad (32)$$

where $\kappa$ is the opacity of plasma. The observed value of Eddington luminosity for a distant observer, however, is

$$L_{ed}^{\infty} = \frac{L_{ed}}{(1+z)^2} = \frac{4\pi GMc}{\kappa(1+z)} \qquad (33)$$

Essentially, $L_{ed}$ corresponds to a critical commoving outward heat flux of

$$q_{ed} = \frac{L_{ed}}{4\pi R^2} = \frac{GM}{\kappa R^2}(1+z) \qquad (34)$$

Using Eqs.(31) and (34), we find that, in the regime of z>>1, the parameter

$$\eta = \frac{q_{trap}}{q_{ed}} \sim \frac{(1+z)}{RM} \qquad (35)$$

grows in a dramatic and an unbounded manner. Therefore sooner or later, there must be a stage when $\eta \to 1$, i.e., $q_{trap} \to q_{ed}$. By the very definition of an ``Eddington Luminosity'', at this stage, the collapse must degenerate into a quasistatic contraction due to trapped radiation pressure[9-14]. Since the luminosity of the collapsing object as seen by a distant observer is $L^{\infty} = -d(Mc^2)/du$, where u is Vaidya time, the time scale associated with this process is

$$u = \frac{Mc^2}{-c^2 dM/du} = \frac{Mc^2}{L_{ed}^{\infty}} = \frac{\kappa c(1+z)}{4\pi G} \qquad (36)$$

Obviously $u \to \infty$ as the BH stage would be arrived, i.e., $z \to \infty$ irrespective of the details of the process. Thus the Eddington limited contracting phase actually becomes eternal and the object in this phase is called an *Eternally Collapsing Object (ECO)*. Therefore, the finite mass BHCs are in the ECO phase though after infinite time, under the assumption of baryon number constancy, they would approach the exact BH state of $z = \infty, ..M = 0$. Atleast *one numerical computation has shown that Newtonian Supermassive stars collapse to form (initially) ECOs rather than BHs[15]*. If the plasma of ECO would be assumed to be a fully ionized hydrogen, one would have $\kappa \approx 0.4 cm^2/g$, the far away ECO luminosity will be

$$L_{\infty} = 1.26 \times 10^{38}(M/M_*)(1+z)^{-1} erg/s \qquad (37)$$

Note actual value of $L_{\infty}$ could be lower by one order because we have taken here the lowest theoretical value of $\kappa$, and the actual value of $\kappa$ could be one order higher. With such a value of $L_{\infty}$, one will have

$$u = 4\times 10^8 (1+z) \text{ yr}. \qquad (38)$$

Since all astrophysical plasma is endowed with some magnetic field and which increases with contraction, it is quite likely that ECOs have strong intrinsic magnetic field and accordingly they are also called ``Magnetospheric ECOs'' or ``MECOs''. A spinning MECO may radiate like pulsar and Robertson & Leiter have developed specific version of MECOs with several crucial assumptions about magnetic properties. In this way, they have tried to show that the X-ray/Radio properties of many BHCs in X-ray binaries as well as quasars may be explained by considering the putative compact object as MECOs rather than featureless true BHs [16,17,18,19]. The present discussion however would be of generic nature independent of specific models.

## 5. General ECO Properties

At high z>>1, any object, including an ECO has a radius practically equal to its Schwarzschild value:

$$R = \frac{2GM}{c^2} \approx 3 \times 10^5 (M/M_*) cm \qquad (39)$$

The mean mass energy density of the ECO is

$$\rho = \frac{3Mc^2}{4\pi R^3} = \frac{3c^8}{32\pi G^3 M^2} \qquad (40)$$

where we have made use of Eq.(39). We earlier showed that any quasistatic object with z>>1, is completely dominated by radiation energy density rather than by rest mass energy density $\rho_r >> \rho_g$ [19]. Consequently the EOS for such an object is $\rho \approx \rho_r = aT^4$ where ``a'' is the radiation constant and T is the mean temperature (this T is not to be confused with Schwarzschild Time). Then it easily follows that the mean temperature of an ECO is

$$T = \left(\frac{3c^8}{32a\pi G^3 M^2}\right)^{1/4} \qquad (41)$$

Numerically, one has

$$T \approx 600(M/M_*)^{-1/2} MeV \qquad (42)$$

Even an ECO with mass as large as M=100 M$_*$ will have a temperature of 60 MeV and it could be sea of *Quark Gluon Plasma (QGP)*. However a supermassive ECO with $M = 10^6 M_*$ will have T ~ 600 KeV, i.e., it indeed be a hot plasma of pairs and baryons though energy density will still be dominated by pairs and radiations rather than by rest mass energy.

## 6. ECO Photosphere

In all hot self-gravitating objects, density must fall off rapidly towards the edge and the mass of the outermost layers should be negligible in comparison to the total mass M. In order that radiation can eventually leak out (howsoever small it may be), the density drop is essential and such an outer region is generally called a ``photosphere''. Because of extreme gravity, the extent of the photosphere also must be negligible in comparison to the radius of ECO. Note that, the ``scale height'' of this region must also decrease as $(1+z)^{-1}$ apart from other factors. Because the redshift in the photosphere is z>>1, it still dominated by radiation energy, its EOS is

$$\rho_p = 3p_p \qquad (43)$$

even though $\rho_p << \rho; .. p_p << p$ (in this section we take G=c=1). Here the subscript ``p'' indicates ``photosphere''. Although the ECO photosphere is not in strict hydrostatic equilibrium (like the photosphere of the Sun or any other hot object), in view of the extremely large time scale of quasistatic balance (Eq. 38), we can safely apply Tolman Oppenhemer Volkoff equation for hydrostatic balance for a quiescent ECO photosphere:

$$\frac{dp_p}{dR} = -\frac{(\rho_p + p_p)(M + 4\pi R^3 p_p)}{R^2(1 - 2M/R)} \qquad (44)$$

Since $M >> 4\pi R^3 p_p$, $\rho_p << \rho; .. p_p << p$, and $\rho_p = 3p_p$, foregoing equation simplifies to

$$\frac{dp_p}{dR} = -\frac{4p_p M}{R^2(1 - 2M/R)} \qquad (45)$$

Recalling that,

$$(1 - 2M/R) = (1+z)^{-2} \tag{46}$$
and,

$$\frac{2M}{R^2} dR = -2(1+z)^{-3} dz \tag{47}$$

we rewrite Eq.(45) as

$$\frac{dp_p}{p_p} = 4 \frac{dz}{1+z} \tag{48}$$

By integrating the foregoing equation, we obtain

$$\frac{p_p}{(1+z)^4} = const \tag{49}$$

Note that the ``constant'' appearing in this equation is *independent of M because there is no M in Eq.(48)*. Since, $p_p = (a/3) T_p^4$ for a radiation energy dominated photosphere, above Eq. leads to

$$\frac{T_p}{1+z} = const \tag{50}$$

Eq.(50) shows that the temperature of the ECO as seen by a distant observer

$$T_\infty = \frac{T_p}{1+z} \tag{51}$$

*is a constant and same for all ECOs! The sum of such blackbody radiation from the ECOs thus should give rise to a universal background thermal radiation.* And this background thermal radiation should mimic the mass distribution of luminous galaxies/matter and will be isotropic if the latter will be so.

Fixing the Value of $T_\infty$

Since the actual value of $\kappa$ could be larger than the lowest theoretical value of 0.4 g/cm² adopted here, let us rewrite Eq.(37) as

$$L_\infty = 1.26 \omega \times 10^{38} (M/M_*)(1+z)^{-1} \, erg/s \tag{52}$$

where $0.1 < \omega < 1.0$.

There is also a general thermodynamic formula for the far away ECO luminosity:

$$L_\infty = 4\pi R^2 \sigma T_p^4 (1+z)^{-2} \tag{53}$$

where $\sigma = 5.65 \times 10^{-5} (cgs)$ is the Stefan-Boltzmann constant. From Eq.(51), we see

$$T_p = (1+z) T_\infty \tag{54}$$

Also, numerically, ECO radius

$$R = 3 \times 10^5 (M/M_*) cm \tag{55}$$

Now using Eqs. (52), (53), (54) and (55), we obtain

$$(1+z) = 1.3 \omega^{1/3} \times 10^{10} (M/M_*)^{-1/3} T_\infty^{-4/3} \tag{56}$$

The above relation self-consistently shows that the ECO photosphere has an extremely high z. Plugging this equation back in (52), we have the ECO luminosity in terms of its observed temperature:

$$L_\infty \approx 10^{28}(M/M_*)^{4/3}T_\infty^{4/3}\omega^{2/3} \text{ erg/s} \tag{57}$$

Thus for the BHC in M87 which has $M \approx 3\times10^9 M_\oplus$, we will have a quiescent bolometric luminosity of
$$L_\infty \approx 5\times10^{40}T_\infty^{4/3}\omega^{2/3} erg/s \tag{58}$$

If one would assume that $T_\infty \sim 1000K$, i.e., infrared, above equation would suggest a large $L_\infty > 10^{46}$ erg/s. But, the quiescent near IR luminosity of M87 is $<10^{41}$ erg/s Such considerations convince us that the value of $T_\infty$ *cannot be much higher than a few Kelvin and one can very well fix it as 2.75 K!* When we do so, we find
$$L_\infty \sim 3\times10^{28}\omega^{2/3}(M/M_*)^{4/3} erg/s \tag{59}$$

The closest supermassive ECO is probably Sgr A$^*$ with $M \sim 10^6 M_*$. In this case, Eq.(59) would predict a total bolometric luminosity of
$$L_{Sgr} \approx 3\omega^{2/3}\times10^{36}M_6^{4/3} erg/s \tag{60}$$
where $M_6 = M\times10^{-6}M$. Note that the microwave radiation from the core of Sgr A$^*$ will however get significantly scattered by dust and electrons. Because of such scatterings, the ``point source'' may appear as a hazy diffused source of much larger angular. *In fact WMAP observations have indeed revealed a ``microwave haze'' of angular width ~ $20^0$ towards the galactic centre.* The estimated luminosity of this ``microwave haze'' is [21]
$$L_{WMAP} \approx (1-5)\times10^{36} erg/s \tag{61}$$
in the band (23-61) GHz. The total bolometric luminosity should be a few times higher and this would be in excellent agreement with Eq.(60) in view of the following ranges of $0.1<\omega<1.0$ and $1<M_6<4$ which is just consistent with observations. Attempts have however been made to link this haze with ``dark matter'' [22]. Considering a distance of d=10 Kpc, the corresponding total energy density at Earth comes out to be

$$u_{Sgr} = \frac{L_{Sgr}}{4\pi d^2 c} \sim 1.6\omega^{2/3}\times10^{-15}M_6^{4/3} erg/cm^3 \tag{62}$$

whereas the energy density of supposed cosmic background radiation is

$$u_{Cbr} = 4.5\times10^{-13} erg/cm^3 \tag{63}$$

Taken at is face value, from the direction of Sgr A, one would then see a fluctuation in the background radiation temperature

$$\frac{\delta T}{T} \sim \frac{1}{4}\frac{\delta u}{u} \approx \omega^{2/3}M_6^{4/3}\times10^{-3} \tag{64}$$

which happens to be of the order of the 'dipole anisotropy'. As the Sun with Earth in tow will move around the galactic centre there could be additional smaller fluctuations in δT. There will also be some contribution to this background radiation due to nearby ECOs (i.e., BHCs). As a result anisotropies in δT are expected along the plane of the ecliptic too. And such anisotropies have also been found by WMAP [22-30].

<u>ECO Luminosity Again:</u>

Let, the photosphere temperature of an ECO be smaller than its mean temperature by a factor of α :
$$T_p = \alpha T \tag{65}$$
Then Eqs.(53) & (56) will give

$$L_\infty = \frac{4\pi\sigma(3\times 10^5)^2 (M/M_*)^2 \alpha^4 T^4}{(1+z)^2} \approx 2\times 10^{58}\alpha^4 (1+z)^{-2} erg/s \tag{66}$$

Further a combination of Eqs. (56), (57) and (61) yield

$$\alpha \approx 0.5\omega^{1/3}\times 10^{-2}(M/M_*)^{1/6} \tag{67}$$

Consequently, from Eqs. (42) and (67), we find that the ECO photosphere is always sufficiently hot.:

$$T_p \approx 3\omega^{1/3}(M/M_*)^{-1/3} MeV \tag{68}$$

The Turbulent Life of a Hot ECO:

The entire picture of ECO depicted above assumed the text book condition of gravitational collapse, i.e. baryon number conservation. Similar assumption is also made for studying, say, solar structure equations. But we know that, in reality, baryons and not radiation alone leave a modestly hot Sun in the form of solar wind and storms. Such radiative activity increases rapidly for hotter stars because of increasing role of radiation pressure. Thus an ECO, being practically supported entirely by radiation pressure is prone to constant radiation driven instabilities. The worst of such instabilities in fact happen during pre-natal stages of ECO formation when radiation driven ECO plasma is thrown out in the form of *Gamma Ray Bursts.* The ECO plasma is almost pure radiation with radiation to baryonic mass density of : $\rho_{rad}/\rho_{rest} \sim z$ [20].

Note the last stage of spherical static configuration corresponds to a gravitational red-shift of z=2.0. Thus GRB must happen when the collapsing object crosses this borderline and attempts to settle down for a large z which could be of the ~ 1000. If there would be no additional baryonic load due to preexisting overlying baryonic layers, then the bulk Lorentz factor of the flinched away QGP plasma would be Γ~z. Since all radiation supported stars are unstable, there could be more of such eruptions before a stellar mass ECO would really settle down (quasistatically) to a range of $z \sim 10^{9-10}$ as indicated by Eq(56). Thus there could be several GRBs from the same direction in the span of a day or may be a year. Such recurring GRBs are indeed seen. Even when the ECO settles down to a very high range of z as indicated by Eq(56)., it is vulnerable to mini bangs driven by uncontrollable radiation instabilities. A quiescent ECO would however synthesize light elements in its envelope (for stellar mass cases) or in its body (for supermassive cases). And such light elements would also be thrown by intermittent ECO flares into the ISM. Light elements apart, ECOs toss out pure QGP or hydrogen into the ISM both during GRBs or during perennial mini-flares. While doing so, an ECO may wither away prematurely after an age which is expected to astronomically significant.

4. Conclusions:

The Schwarzschild solution is indeed correct and exact; but the integration constant for the vacuum solution is zero and hence the *observed* BHCs cannot be true BHs with $z = \infty$. On the other hand, they must be hot quasistatic ECOs with $\infty > z \gg 1$. Under the assumption of baryon number conservation, an ECO would asymptotically approach the ideal state of a true BH with M=0 and z=∞. Very massive/dense objects undergo continued gravitational collapse to become ECOs and their formation is marked by occurrence of GRBs. All ECOs cook light elements at appropriate cool regions having T<~1 MeV and spew off the same in the ISM during incessant radiation driven flares. Most importantly, the observed black body temperature ECO photospheres has a unique value of few Kelvins. And thus the observed thermal microwave background radiation could be due to cosmic contribution of ECOs. If such an assumption is made, there is good matching between the predicted and observed values of the microwave luminosity of Sgr A$^*$.

According to this prediction, all galactic ECOs, i.e., BHCs would modulate the supposed ``primordial radiation field''. Thus as Sun would move around the galactic centre with Earth in tow, various kinds of fluctuations, asymmetries and anisotropies in this back ground radiation should be seen. And indeed various unexpected and unexplained anisotropies are seen for this background radiation [23-31].

**Acknowledgement**: The author thanks Stanley Robertson for a careful reading of the manuscript and useful discussions. The organizers are also thanked for offering partial financial help.